\documentclass[journal]{IEEEtran}
\usepackage{graphicx,cite,subfigure}

\usepackage[colorlinks]{hyperref}

\usepackage{amsmath}

\usepackage{footnote}

\begin{document}
%
\title{A Vision to Smart Radio Environment:\\Surface Wave Communication Superhighways}

\author{Kai-Kit Wong,~\IEEEmembership{Fellow,~IEEE,}
        Kin-Fai Tong,~\IEEEmembership{Senior Member,~IEEE,}
        Zhiyuan Chu, 
        and Yangyang Zhang

\thanks{K. Wong, K. Tong, and Z. Chu are with the Department of Electronic and Electrical Engineering, University College London, London WC1E 7JE, UK.}
\thanks{Y. Zhang is with Kuang-Chi Institute of Advanced Technology, China.}
\thanks{Manuscript received April XX, 2020.}}

%
%

\markboth{Submitted to IEEE XXXX}%
{6G Wireless System}
%



\maketitle

\begin{abstract}
Complementary to traditional approaches that focus on transceiver design for bringing the best out of unstable, lossy fading channels, one radical development in wireless communications that has recently emerged is to pursue a smart radio environment by using software-defined materials or programmable metasurfaces for establishing favourable propagation conditions. This article portraits a vision of communication superhighways enabled by surface wave (SW) propagation on ``smart surfaces'' for future smart radio environments. The concept differs from the mainstream efforts of using passive elements on a large surface for bouncing off radio waves intelligently towards intended user terminals. In this vision, energy efficiency will be ultra-high, due to much less pathloss compared to free space propagation, and the fact that SW is inherently confined to the smart surface not only greatly simplifies the task of interference management, but also makes possible exceptionally localized high-speed interference-free data access. We shall outline the opportunities and associated challenges arisen from the SW paradigm. We shall also attempt to shed light on several key enabling technologies that make this realizable. One important technology which will be discussed is a software-controlled fluidic waveguiding architecture that permits dynamic creation of high-throughput data highways.
\end{abstract}


%
\IEEEpeerreviewmaketitle

\section{Introduction}
Wireless communications has come a long way, from mobile telephony in the 70s to nowadays massive, dynamic multimedia information access anywhere anytime in the era of Internet of everything (IoE). We have also witnessed a shift in research focus from outdoors to indoors since most data demand now tends to take place in indoor environments. This has motivated the emerging concept of smart radio environment which uses software-defined materials or software-controlled metasurfaces \cite{Akyildiz-18,Marco-19} to engineer a radio environment suitable for wireless communications. Many believe this will play a role in 6G \cite{Tariq-20}.

The use of software-controlled metasurfaces for improving wireless communications is a thriving research area. A majority of recent efforts have been devoted to the study of adopting passive radiation elements on a programmable surface that has the ability to apply arbitrary phase shifts on the receiving radio signals and reflecting them constructively as a focussed beam towards the intended receiver. This approach is widely known as reconfigurable intelligent surfaces (RISs) \cite{Ntontin-20}.

RIS is particularly attractive for their low power consumption and hardware cost, because of the use of relatively cheap radiating elements. It can be interpreted as using large surfaces present in the environment as a large aperture for collecting and transmitting radio signals for improved energy efficiency. Remarkably, it was reported in \cite{Dai-19} that it is possible to achieve a 21.7 dBi antenna gain using an RIS with 256 2-bit elements at 2.3 GHz, and a 19.1 dBi antenna gain at 28.5 GHz. It should be noted that programmable metasurfaces can also be used to directly modulate radio-frequency (RF) carrier signals, without the need of standard components such as mixers, filters, and phase shifters, greatly simplifying the hardware complexity for wireless communications systems \cite{Tang-19}.

The concept of smart radio environment is, however, much more than a low-cost alternative to relaying, beamforming and communication transceivers, and represents a new paradigm of engineering the radio environment through carefully designed, software-controlled metamaterials (or ``meta-atoms'') that can alter their electromagnetic (EM) properties to suit the purpose of various communication applications. Reducing interference, enhancing security, and extending the range of communication are amongst the most obvious applications \cite{Akyildiz-18}. Although the main advantages of metasurfaces are their low hardware cost and power consumption, such as in the case of RIS, utilizing programmable metasurfaces to create a smart radio environment may mean that additional signal processing and network intelligence will add to the cost and power consumption.

This article proposes a new vision of smart radio environment that considers the use of {\em non-radiative, trapped surface wave propagation} \cite{Barlow-1953,Tong-2019}, as opposed to free-space propagation where radio waves are launched from the surface in \cite{Akyildiz-18,Marco-19}. 
The surface waves considered in this article are trapped surface waves \cite{Tong-2019} which glide at the interface of materials with different dielectric constants and the radio propagation is made to be confined to the surface. A unique advantage of surface wave communications (SWC) over free-space communications (FSC) is its much more favourable pathloss, which is inverse of the distance $1/d$, instead of the inverse of the squared distance $1/d^2$ in the case of FSC. Also, confining the communication to the surface means that interference management becomes a lot easier since communication can be managed on a particular pathway using software-controlled waveguiding surfaces, the concept that can be enabled by a software-controlled fluidic structure \cite{Fortuny-17}. The outcome resembles a transportation network of SWC superhighways on surfaces of meta-atoms, providing various functionalities of a smart radio environment.

The rest of this article is organized as follows. In Section~\ref{sec_Barlow}, we provide a high-level background of SWC and highlight the unique advantages that make it particularly appealing for the smart radio environment application. Section \ref{sec_vision} presents our vision of SWC superhighways. Then Section \ref{sec_enablers} describes the key enabling technologies for software-controlled SWC while Section \ref{sec_chall} discusses the main challenges of the proposed SWC paradigm. Finally, we conclude this article in Section \ref{sec_con}.

\section{Surface Wave}\label{sec_Barlow}
\begin{figure}[]
\begin{center}
\includegraphics[width=0.95\linewidth]{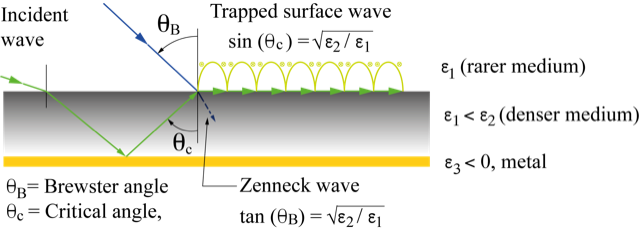}
\caption{Illustration of different angles and the corresponding waves.}\label{fig:sw0}
\end{center}
\end{figure}

Surface wave is a non-radiating wave that propagates along the interface between two different media \cite{Barlow-1953}. The definition can be formally classified into eleven types according to their physical properties \cite{Schelkunoff-59}. When a radio wave is incident at a boundary from a denser medium and if the incident angle is equal to or greater than the critical angle of the media, then the radio wave will be `trapped' in the denser medium, with the evanescent fields in the rarer medium, and the wave will be confined to the surface. Figure \ref{fig:sw0} illustrates the geometry of the directions of the waves at the interface. In practice, both media have finite losses, and the E-field of the surface wave will attenuate as it propagates along the interface.  A classical result is that the power of a trapped surface wave is inversely proportional to the propagation distance, $d$ \cite{Barlow-1953}:
\begin{equation}
P_{\sf SWC}(d)\propto\frac{1}{d},
\end{equation}
which is much more desirable than the inverse squared law of what normally occurs in space wave propagation or FSC.

The surface resistance is associated with energy dissipation and determines the attenuation of the surface wave in the propagation direction whereas the surface reactance is associated with the energy stored in the interface, which defines the decay of the wave away from the surface in the propagation direction. The higher the surface reactance, the more closely the energy is stored to the surface and hence the more closely bound the wave is to the surface. Rectangular apertures of finite height are usually adopted as transducers to excite surface wave, with minimal space waves and reflected waves.

\begin{figure}[]
\begin{center}
\includegraphics[width=0.8\linewidth]{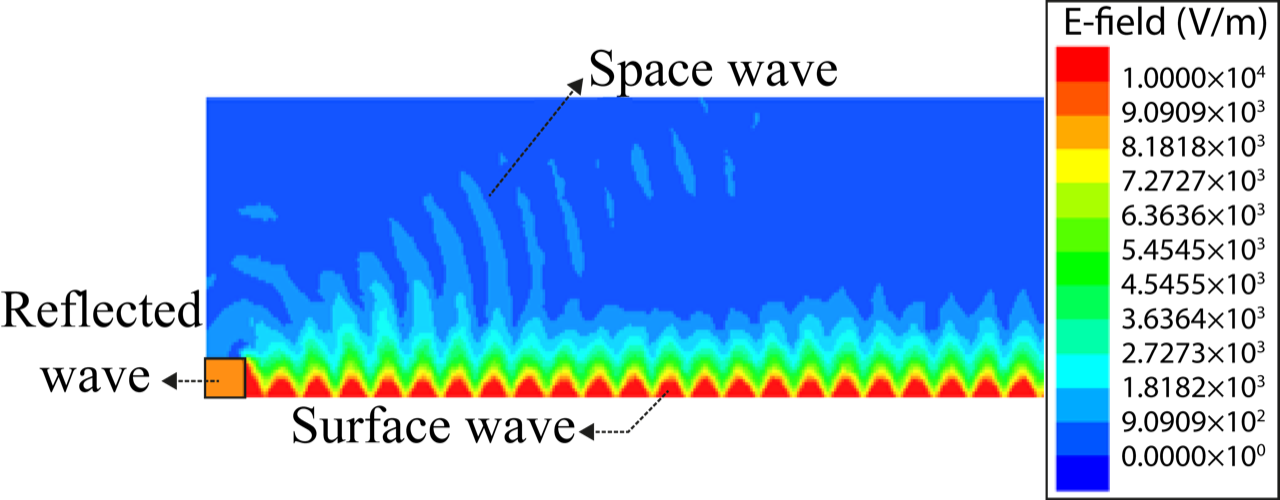}
\caption{Simulation results showing the E-field distribution along the direction of propagation at 60 GHz with the surface impedance of $j200\Omega$.}\label{fig:sw2}
\end{center}
\end{figure}

The most effective surface for SWC is the one which has a purely reactive surface impedance. Two possible approaches to obtain high impedance surfaces are corrugated surfaces and dielectric coated conductors. Corrugated surfaces have purely reactive impedance that depends upon the dimensions of the grooves and humps of the surface. The main limitation of a corrugated structure is its directional periodic structure, which is difficult to fabricate in millimeter-wave frequency bands. A more viable solution is the use of dielectric coated conductors which are much easier to make. The dielectric layer should have a high dielectric constant and low conductivity. Also, the surface impedance can be further adjusted by layering several different dielectric layers on top of each other. In \cite{Tong-2019}, a 52 GHz wideband trapped SWC system by utilizing the dielectric coated conductor approach was implemented. Figure \ref{fig:sw2} shows the E-field distribution along a dielectric coated conductor with surface impedance $j200\Omega$ at 60 GHz.

\section{The Vision: SWC Superhighways}\label{sec_vision}
\begin{figure}[]
\includegraphics[width=1\linewidth]{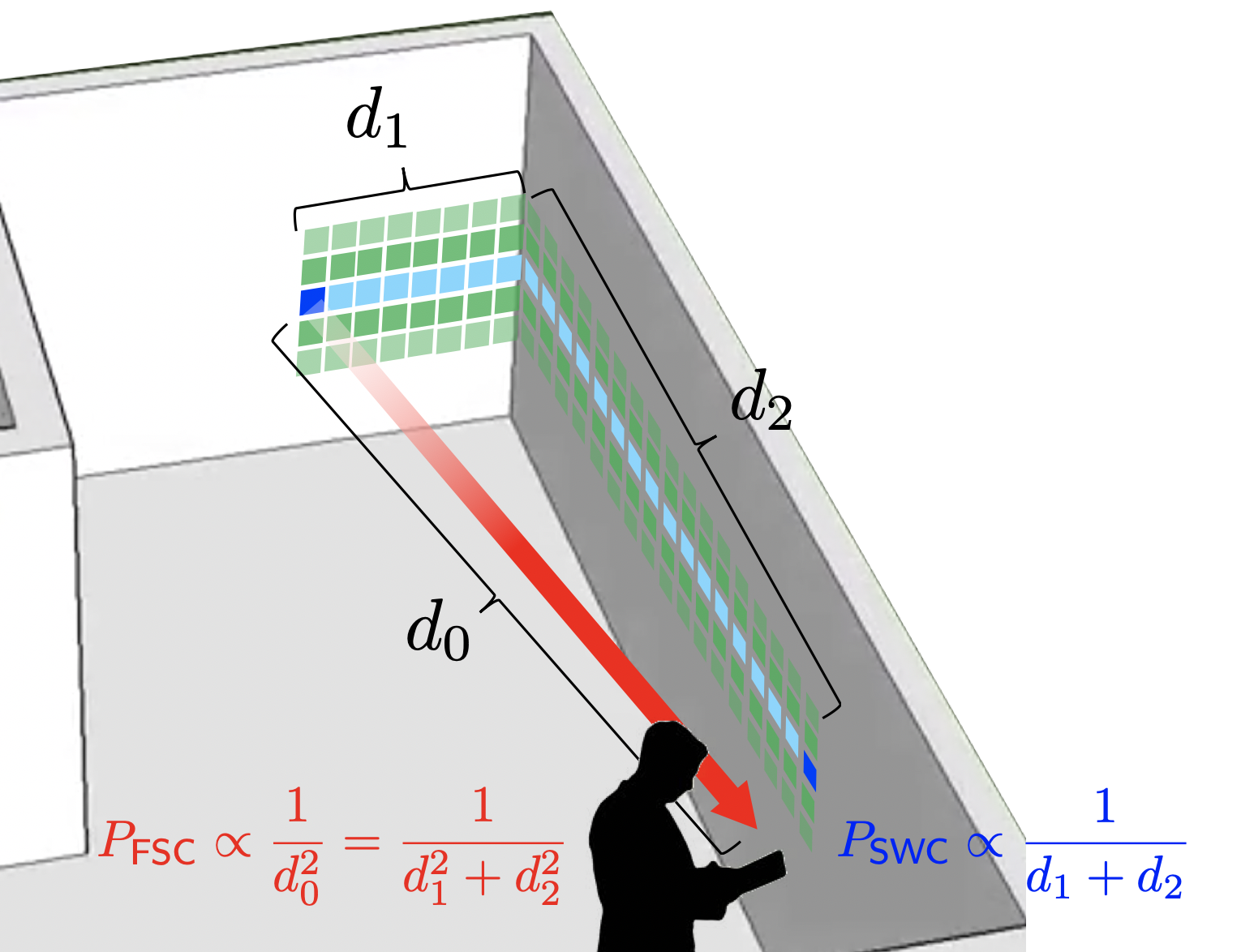}
\caption{SWC versus FSC: Longer distance beats shorter distance.}\label{fig:pathloss}
\end{figure}

Less propagation loss of SWC means that it can be preferable to take a detour and travel a longer distance along walls or surfaces but still have higher received power than a direct path propagating in the free space, see Figure \ref{fig:pathloss}. SWC is thus super-energy-efficient and data can  reach farther distance with the same energy consumption. More remarkably, confining the communication signals to the surface helps contain high-speed data streams, and simplifies interference management.

\begin{figure*}[]
\begin{center}
\includegraphics[width=1\linewidth]{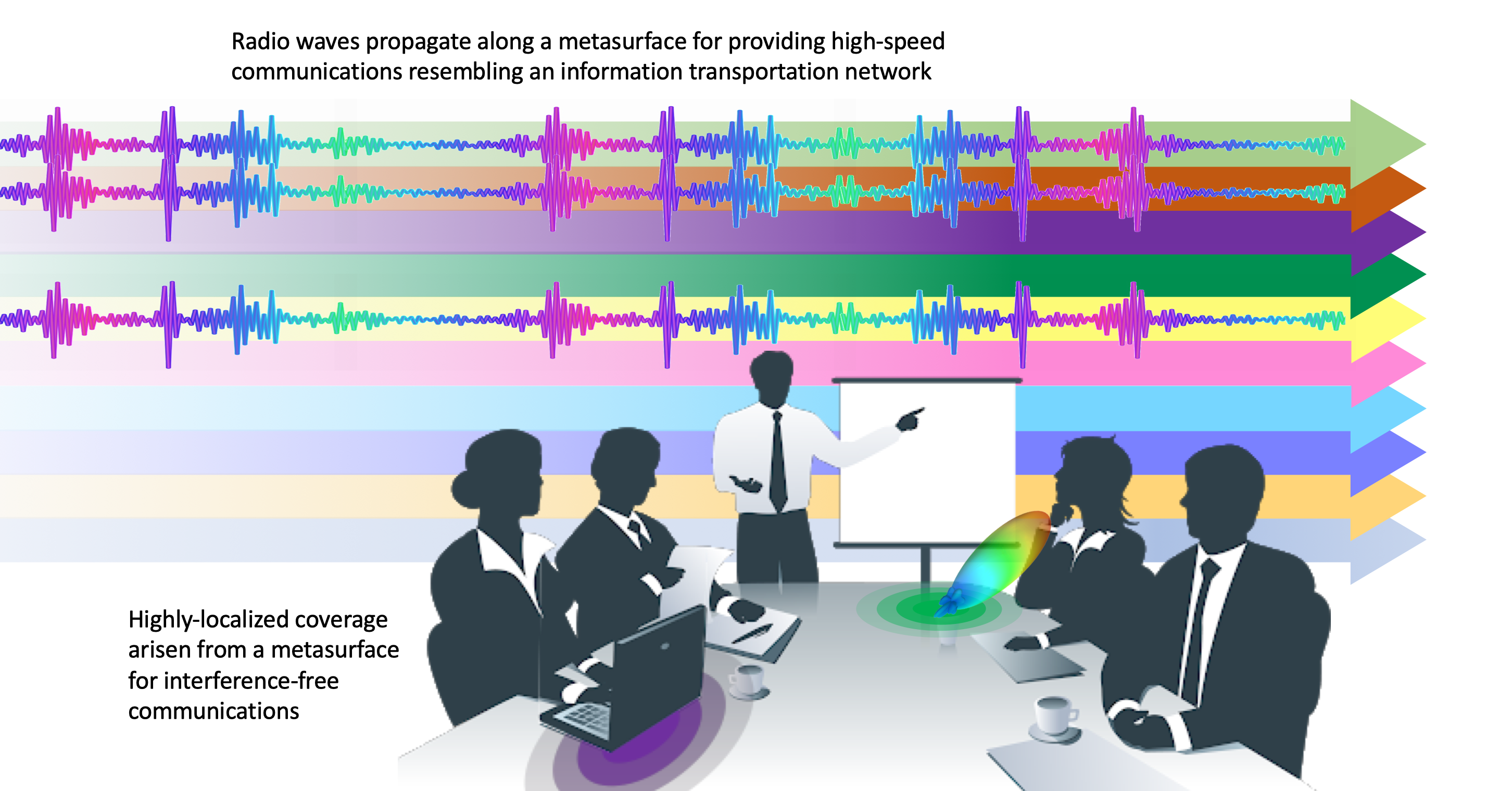}
\end{center}
\caption{The vision of SWC superhighways in a smart radio environment with exceptionally localized wireless coverage (contactless or not).}\label{fig:vision}
\end{figure*}

Our vision of the future smart radio environment therefore resembles a transportation network of communications superhighways with exceptionally localized data access, as depicted in Figure \ref{fig:vision}. In this vision, dedicated pathways are created on programmable metasurfaces to carry superfast data that travel along the surfaces to reach the user equipment (UE) or arrive at a position near the UE. In the latter case, FSC will provide the last hop from the metasurface to the UE over a very short distance. The pathways are software-controlled to adapt to the radio environment and always provide the best routes requiring the least power consumption to the UEs free of interference, thanks to the extraordinary spatial selectivity of SWC.

\begin{figure}[]
\includegraphics[width=1\linewidth]{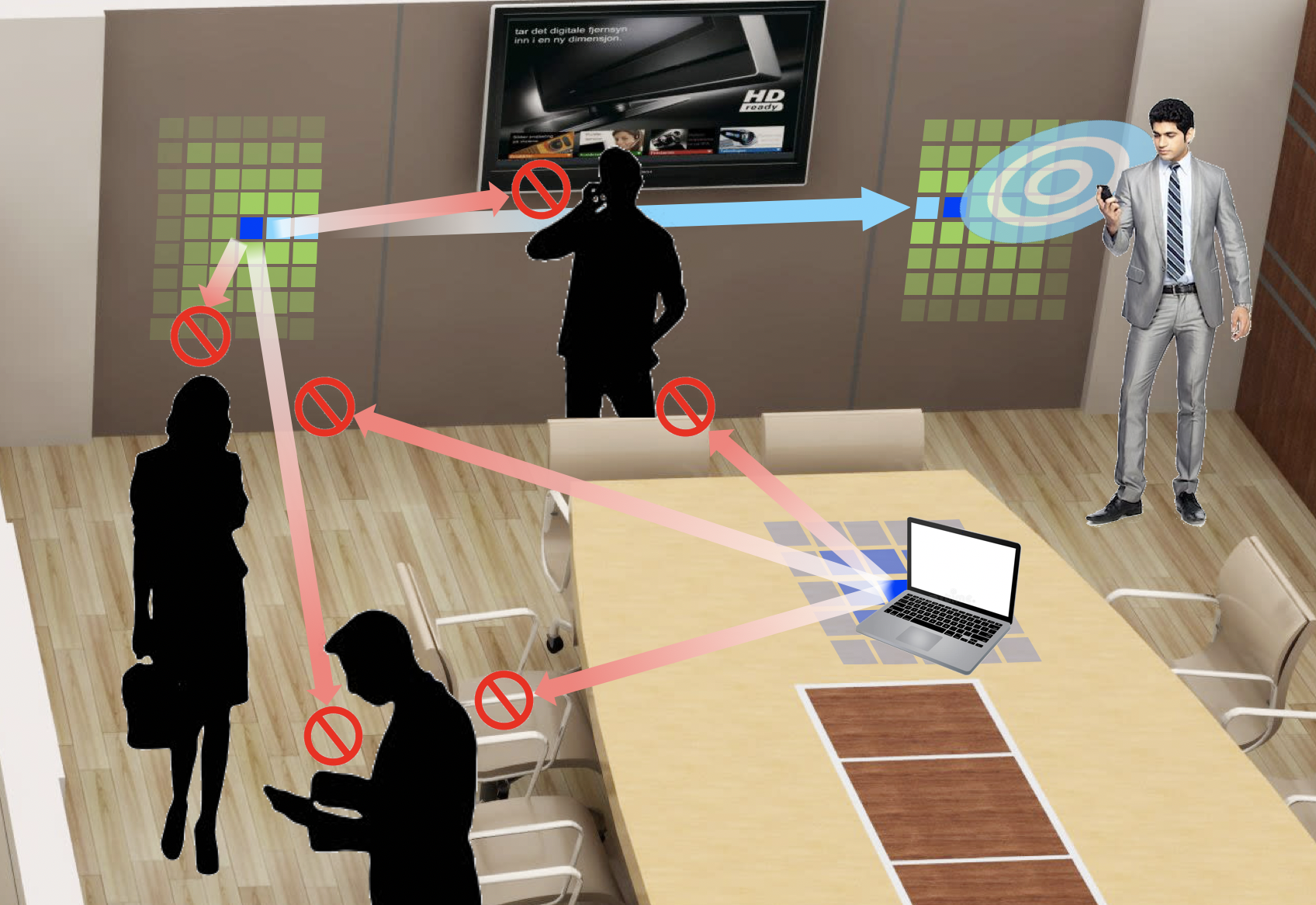}
\caption{SWC inherently confines communication signals to the surfaces and causes no interference to coexisting UEs without the need of sophisticated signal processing and interference management.}\label{fig:int}
\end{figure}

In this vision, one reality is that radio signals only appear where they should, and wireless coverage, contactless or not, is exceptionally localized. By contrast, communications relying on FSC tends to have radio waves unintentionally occupying the entire space unless multiple antennas are used to beamform the radio signals to be confined into certain space that requires advanced signal processing and resource management. This is the inherent problem in wireless communications. As a matter of fact, much processing and intelligence in wireless networks for 5G and beyond go to the management and control of radio signals for massive connectivity that allow signals to coexist without causing harmful interference to each other. This is a very challenging task because radio waves naturally propagate in all directions, and when radio waves hit objects along the way in the environment, the reflection and diffraction further complicate the interference pattern. This task will, however, be greatly simplified in SWC, since the radio waves will be kept on the surface and their presence is absolutely predictable. As envisaged in Figure \ref{fig:int}, in addition to walls, surfaces like office desk can also be equipped with meta-atoms to provide zonal, targeted data access, making possible the ultimate interference-free communications in indoor environments.

A natural result of this vision is that interference management becomes traffic control, and it is highly location-driven. In other words, the key will be to localize all the UEs relative to the infrastructure of metasurfaces in the environment, and determine the best possible routes to serve the UEs. Interestingly, localization in the SWC paradigm can be less complex. The reason is that there will be a large number of meta-atoms as anchors with known locations available in the environment that can take part in the localization task for UEs. In addition, the UEs are likely to be in close proximity of the anchors, and have direct line-of-sights (LoS) for ranging and localization. The assurance of sufficiently large number of direct LoS paths within a short distance from the UEs makes low-complexity high-resolution localization realizable. It is also worth pointing out that simultaneous SWC in the same pathway is possible as the metasurface can operate on a wide frequency band. 

Another feature in the SWC-based smart radio environment paradigm is the massive deployment of antennas on surfaces. Apart from the cases where communications takes places via direct contact with the UEs (e.g., laptops on desk), as in the conventional RIS concept, the surfaces will be equipped with antennas as widely as possible so that the UEs can be reached using short-range FSC anywhere in the environment. The key difference is that the meta-atoms on the surfaces now need to switch between FSC and SWC, by acting as radiating elements and the propagation medium, respectively, depending on the situations, and such switching is done seamlessly.

\section{Enabling Technologies}\label{sec_enablers}
The vision of the SWC paradigm is not a dream but can be a realistic revolution being sought in 6G. In this section, we offer some ideas in the potential enabling technologies.

\subsection{Dual-Functionality Meta-Atoms}
The SWC paradigm suggests a hybrid communication network taking full advantage of SWC and FSC. To realize this, each meta-atom may serve as a tuneable propagation medium or radiating element at any given time and needs to be able to switch between the two functionalities. SWC propagation over a metasurface as the denser medium is a natural phenomenon, and the tricky part is how the signal comes off the surface to reach the intended UE. In general, two `exits' are possible. In the first type of `exit' where the UE is in direct contact with the metasurface (for example, a laptop on a ``smart'' desk), it is rather straightforward to have a transducer integrated on the UE to easily capture the signal and receive the data.

By contrast, in the case where the UE is not in contact with the surface, it requires that the metasurface have the capability to transform SWC into space wave to propagate over air to leap to the UE. Also, it is expected that the metasurface when acting as radiating elements, has the signal processing capability to form focused signal beams towards the intended UE and avoid interference, in the same way as in the RIS applications.

In the leaky-wave and holographic antenna literature, periodic metallic geometries have been well studied to radiate an RF signal from a surface \cite{Antarr-08}. In particular, a trapped surface wave can be diffracted by a periodic metallic geometry that will let part of the surface wave scatter into the free space as space wave. It has also proven possible to accurately control the direction of the radiation off the metasurface \cite{Johnston-10}. Several approaches have been proposed to reconfigure the amount of radiation and the angle of departure of the space wave from a diffracted surface, and they include using active semiconductor devices and metamaterials \cite{He-19}. Furthermore, recent advances in transparent conductive sheet using graphene, carbon nanotube or metallic compounds will help the realization of meta-atoms that can be lightweight and invisible on the surfaces.

\subsection{Software-Controlled Fluidic Waveguiding Structures}\label{sec_enablers_B}
The SWC vision is only realizable if we have a mechanism to create on demand dynamic pathways on surfaces for SWC. This is important for interference management and pathloss reduction of the data streams on the metasurface. Normally, an RF signal that leaves the surface wave transducer will travel along the surface wave plane, following a radial pattern with an angular coverage determined by the electrical size of aperture width of the transducer \cite{Tong-2019}. Research on dynamic creation of pathways on the surface is very limited although high surface impedance will surely enhance surface wave propagation. A great deal of research efforts in the SWC area have been trying to realize reconfigurable surface impedances on a surface.

\begin{figure*}%
\centering
\subfigure[Tubes on surface, with those in blue filled with conductive fluid.]{%
\label{fig:6a}%
\includegraphics[height=1.8in]{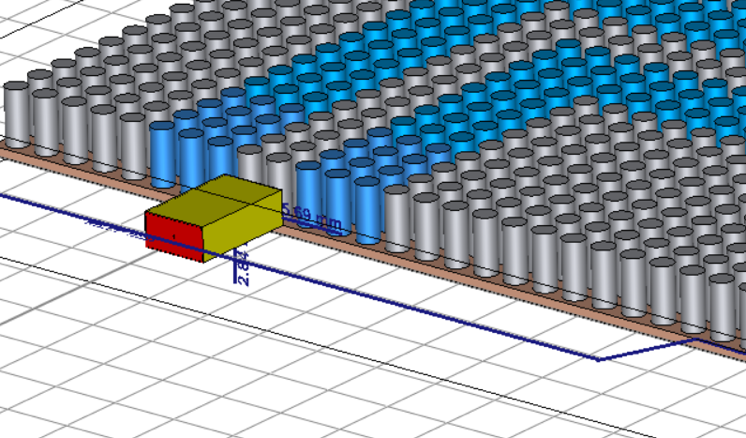}}%
\qquad
\subfigure[Guided surface wave propagation over a preset pathway.]{%
\label{fig:6b}%
\includegraphics[height=2in]{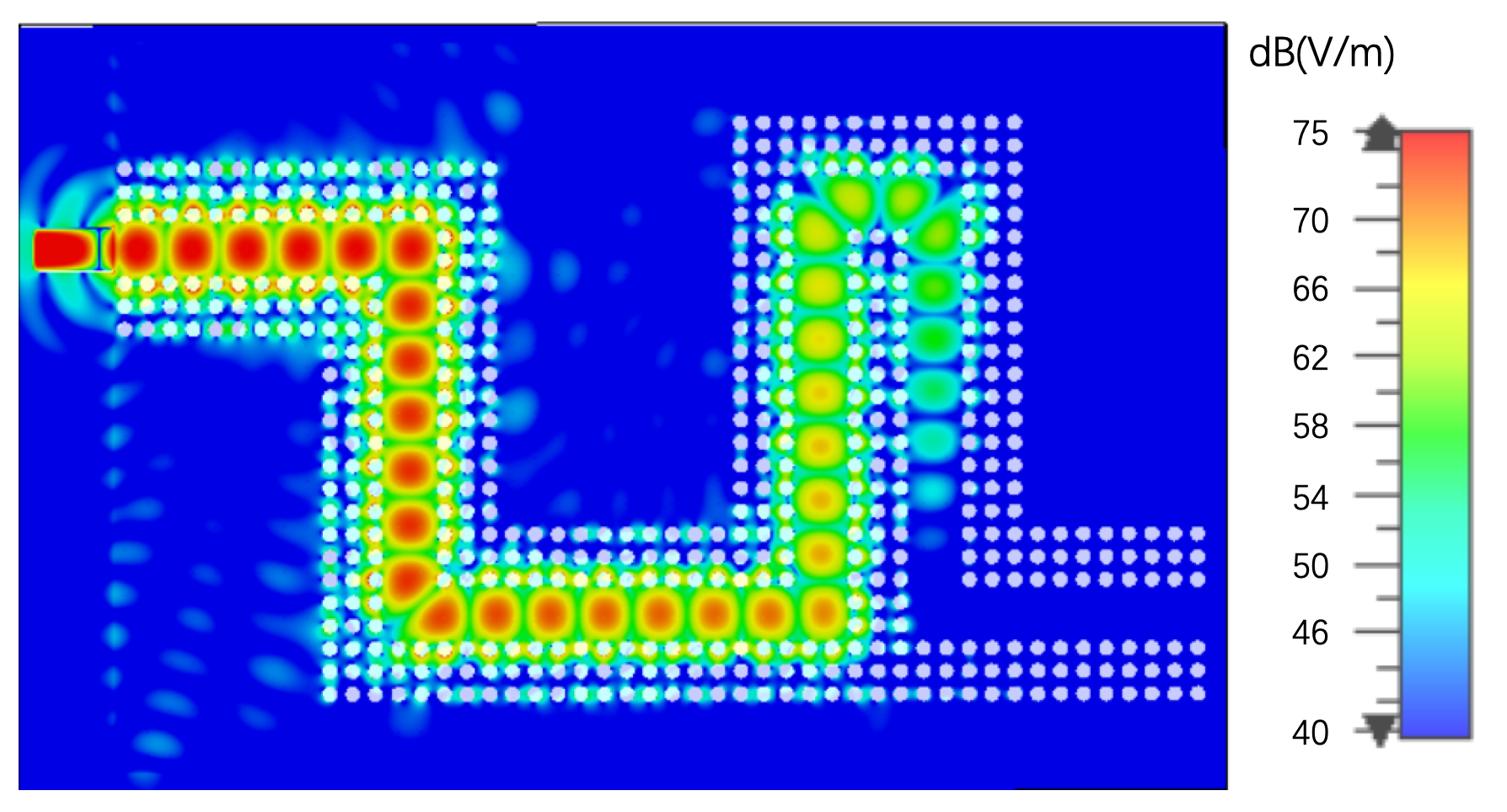}}%
\caption{Illustration of the software-controlled waveguiding structures for dynamic pathway creation with surface impedance $j96.5\Omega$ at 28 GHz.}\label{fig:fluid}
\end{figure*}

Flexibly and dynamically creating pathways on a surface is a much more challenging task but practically achievable. One possible approach is to leverage a microfluidic system where a large number of micro-tubes are pre-fabricated in the surface substrate of a few millimeters thickness. The micro-tubes are connected to an array of software-controlled pumps which can inject conductive fluid into the tubes if required. If some of the pumps are activated, a selected group of the micro-tubes will be filled with conductive fluid, which then creates an integrated waveguiding structure to form a tunnel of pathway. The pattern which the conductive fluid-filled micro-tubes form, determines the pathway for SWC. In Figure \ref{fig:fluid}, some preliminary results are provided to illustrate the feasibility of such concept with a surface impedance of $j96.5\Omega$ at the operating frequency of $28$ GHz. In this example, Galinstan that is a liquid metal alloy with high conductivity is chosen as the conductive fluid.

It is worth noting that software-controlled fluidic structures have recently been investigated for antenna design \cite{Fortuny-17}. The knowhow in that application is anticipated to be useful in the engineering and signal processing of the microfluidic system for programmable metasurfaces. This architecture makes possible joint optimization of the signal and data streams, the resource allocation and management of the communication, and the propagation media that accommodate the communications. Intelligence in such holistic approach will be essential.

\subsection{Artificial Intelligence (AI) Empowered SWC}
Recent advances have seen AI especially deep learning to be given a major role for 5G and beyond communications \cite{Tariq-20}. There have been numerous successful examples of employing AI for wireless communications, from physical layer designs such as channel estimation and precoding, to network resource allocation such as traffic control and user pairing, to security and authentication, and to topology formation and management, to fault prediction and detection, and so on. Parallel can be drawn in the SWC paradigm and AI will serve as the brain to empower the superhighway network on the metasurface.

There are several technical directions in which AI is anticipated to be key to the realization of the concept. One obvious avenue is the localization of UEs in the smart radio environment where connections need to be established seamlessly. In the SWC paradigm, the UEs' locations are most important as most communication remains on the surface but only leaps to reach the UEs in the last hop if required. The availability of the locations of the UEs allows the data traffic to be managed on the surface with carefully designed pathways such that the power consumption (and hence the pathloss) is minimized and the interference over the surface is also eliminated.

Localization of the UEs is expected to be done easily from simple energy sensing, as the last hop from the surface to the UEs is likely to be short-ranged and has the direct LoS. Also, the large number of meta-atoms ensures more than sufficient LoS paths to exist and localize the UEs in a way similar to the traditional multilateration approach. The difference is that in this case, most of the paths come from the same direction with different energy levels, instead of paths coming from a variety of directions. Apart from this, metasurfaces are preinstalled to form a smart radio environment, meaning that a fingerprinting approach can be effectively utilized to localize the UEs. This can be addressed using deep learning by training an artificial neural network (ANN) over some simulated energy level data given the floor plan of the environment. Real-time localization of the UEs can then be easily achieved by a simple pass to the ANN after taking energy measurement from the meta-atoms.

AI can also be useful to find the best pathways for the data to travel from the base station (which is now equipped with a surface wave launcher connecting to a metasurface network) to the data-receiving UEs. Such SWC map can be derived from simulated data using deep learning. The key would be to link the UEs' locations and data traffic demand with the physical structures and resources of the metasurfaces making up the radio environment. The optimal SWC superhighway network as a function of the UEs' parameters will be learned.


\section{Opportunities and Challenges}\label{sec_chall}
\subsection{Fully Reconfigurable Meta-Atoms}
Metamaterials have been researched for nearly two decades, having generated many mind-blowing results including cloaking that can make objects invisible to sensors by controlling their EM radiation. For mobile communications, metamaterial-based antennas are also hopeful to make small-sized highly efficient wideband antennas possible. The notion of metasurface-empowered smart radio environments is expected to make a huge impact in 6G and has been led by the mainstream efforts of RIS where passive meta-atoms are considered. Despite the promising results of RIS, the fact that the meta-atom is based on the current microstrip patch antenna technology means that the bandwidth unfortunately tends to be narrow, which hinders its development for high frequency bands. Exploring the use of double negative (DNG) metamaterials would be key to unlock the bandwidth limitation and miniaturize the meta-atoms for increasing the aperture for performance improvement.

In addition, the SWC paradigm requires that the meta-atoms not only act as antennas for radiation but operate as a medium for surface wave propagation. For the latter mission, research is required to devise a technology to adaptively control the characteristic impedance of the meta-atoms so that the surface can be optimized for adhering the radio waves to the surface for superior propagation loss and ease of interference management. A micro-electronics approach that can achieve fine control and resolution of the  impedance of a meta-atom via a digital signal should be studied. Note that the current state-of-the-art meta-atom technology nevertheless comes with the limitation that the amplitude and phase of the radiation off the meta-atoms are strongly coupled. This will need to be tackled if metasurfaces are to be fully intelligent for advanced coding and signal processing. Even if this can be achieved, deciding on the appropriate surface impedance is far from trivial which couples with the action to create SWC pathways that affects the characteristic impedance of the medium above the surface.

Depending upon the communication networking and whereabouts of the UEs, it may also be necessary for a meta-atom to act as a radiating element on one frequency but a propagation medium for SWC on another frequency at the same time. Such dual functionality will certainly need a more advanced meta-atom technology that can achieve independent control of the radiation characteristics over multiple frequency bands. All in all,  programmable, reconfigurable and multi-functional meta-atom technologies will need to be sought. 

\subsection{Pathway Creation and Control}
Creating dynamic pathways on surfaces is the central idea of the SWC paradigm. In Section \ref{sec_enablers_B}, an enabling technology which utilizes a microfluidic system of liquid metal to provide on-demand pathways is discussed. However, many obstacles are yet to be overcome. One apparent issue is the choice of the conductive fluid used for creating the pathways. Liquid metals such as mercury are toxic and should not be used. There is also a nontrivial trade-off between the response time which is related to its density and the conductivity of the fluid. More will need to be investigated to fully understand the practicality and feasibility of such approach in a living environment.

Another pressing challenge is the fabrication of tube spaces in millimetre scale for creating an adaptive metasurface. It gets much more difficult when realizing the fact that micro-tubes do have finite thickness and will distort propagation along the fluid-made pathways. A thorough analysis and proper design of the architecture that integrates with the micro-fluidic system to make possible rapid distribution of conductive fluid with great precision and control will be indispensable.

Moreover, the size of the micro-tubes and their spacing are frequency-dependent. How to have an implementable structure that permits flexible size and spacing to accommodate different frequency bands is extremely challenging, not to mention the difficulty of realizing different pathways in the same space at the same time for different frequencies. Although this issue may be mitigated by careful frequency planning and pathway optimization, much more research will need to be conducted to obtain a feasible architecture for dynamic waveguide technologies such as the fluid-based approach discussed above.

\subsection{Model-Driven AI Signal Processing and Networking}
Signal processing and communication networking in such a smart radio environment depend greatly upon the floor plan of the indoor environment because metasurfaces on walls dictate the paths and positions in which data can be delivered to the UEs. One would expect that it is possible to use the floor plan as a model to develop and train a deep ANN that takes the UE's locations and service demands as inputs and produces as output the signal processing and networking solution. If such AI solution becomes available, then real-time optimization of the signal processing for SWC will be straightforward. Doing this, however, requires several obstacles to be tackled.

The first hurdle is to come up with a general representation of the floor plan that contains the essential information in a concise and manageable fashion. This task will impact on the training efficiency of the ANN that extracts the logical features of the environment and translates them into parameters that are ready for the optimization. Secondly, the inputs to the ANN need to be properly defined which may be the locations of the UEs,  their service demands, or even their energy signatures. It is in fact more challenging to define the outputs of the ANN as the variable space can be gigantic. This could include, from the coding rate of transmission to the transmit power, the pathway, the frequency allocation, the beamforming of meta-atoms to even security-related processing and remote charging. Besides, the physical constraints of the meta-atoms and the metasurface as a whole will need to be incorporated in the design.

Even after the inputs and outputs of the ANN are defined, it is unclear how the ANN can be trained properly. Supervised learning requires the availability of labelled training examples whereas unsupervised learning looks for undetected patterns of datasets with no pre-existing labels. There is  a long way to figure out how reliable datasets can be obtained to train such ANN. Ad hoc heuristics by using a combination of traditional techniques should present useful examples to start building a good dataset. Transfer learning is also expected to be useful to ensure generalizability to cope with a range of situations.

\subsection{User-Centric Context-Aware Service-Based Coverage}
The concept of smart radio environment suggests strongly an integration between UEs and the environment. The large area of metasurfaces makes available multi-dimensional time-series of energy spectrum of the UEs that captures very rich behavioural data of the users and contextual information. The enriched understanding of the UEs and their needs mean that context-aware service and applications can be provided to the level that has not been achievable before. Note that the large number of meta-atoms brings the service of a massive number of sensors as well that enable numerous applications such as remote patient monitoring. Opportunities are plenty in terms of applications but it would be tricky to interpret the signals from sensors with highly uneven distributions.

\subsection{Security and Anti-Wiretapping}
With increasing reliance on mobile applications on our daily life, security has become a major concern. In 5G and beyond systems, physical-layer security schemes appear to provide an additional layer of defence to complement traditional cryptographic approaches by exploiting the 
randomness of wireless channels and their unreplicability. The SWC network however makes the communication data more exposed since the signals glide on surfaces and are more predictable, although one merit of SWC is its high predictability to allow simple interference management. Wiretapping on metasurfaces is a real threat that needs to be looked at carefully and addressed. 

Anti-wiretapping techniques will be key to ensure security on metasurfaces. Apart from operating as radiating and propagation elements, meta-atoms should act as sensors to probe, identify and localize any suspected wiretappers or adversaries so that SWC can be rerouted. Together with AI, metasurfaces should possess the brain power to predict suspicious activities of malicious intrusion, and optimize SWC accordingly. 

\section{Conclusion}\label{sec_con}
Recent research has seen the notion of smart radio environment emerge as a mainstream effort to shape the environment to support communication needs by using software-controlled metasurfaces. Contrary to the conventional studies, this article has advocated the vision to utilize metasurface not only as a radiating platform but propagation medium taking advantage of SWC for much less pathloss and simple interference control. The SWC paradigm greatly reduces unnecessary radiation off the surfaces and only beams to the UEs in the last leg if needed. This novel SWC concept is made possible by several enabling technologies, including the one that can dynamically adapt communication pathways on a metasurface by digitally controlling a microfluidic system of liquid metal, all of which have been discussed. This article has also touched upon the opportunities and challenges that come with the vision. It is hoped that this article will serve as a catalyst to trigger further research to make the SWC vision practically feasible.

\ifCLASSOPTIONcaptionsoff
  \newpage
\fi


%

\begin{IEEEbiography}[{\includegraphics[width=1in,height=1.25in,clip,keepaspectratio]{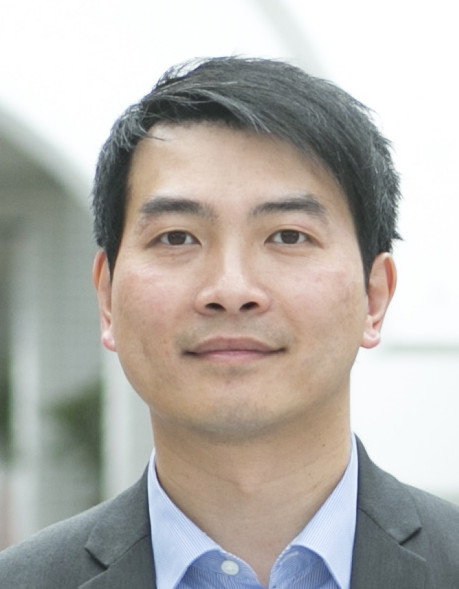}}]{(Kit) Kai-Kit Wong} (M'01-SM'08-F'16) received the BEng, the MPhil, and the PhD degrees, all in Electrical and Electronic Engineering, from the Hong Kong University of Science and Technology, Hong Kong, in 1996, 1998, and 2001, respectively. After graduation, he took up academic and research positions at the University of Hong Kong, Lucent Technologies, Bell-Labs, Holmdel, the Smart Antennas Research Group of Stanford University, and the University of Hull, UK. He is Chair in Wireless Communications at the Department of Electronic and Electrical Engineering, University College London, UK. 
 
His current research centers around 5G and beyond mobile communications. He is a co-recipient of the 2013 IEEE Signal Processing Letters Best Paper Award and the 2000 IEEE VTS Japan Chapter Award at the IEEE Vehicular Technology Conference in Japan in 2000, and a few other international best paper awards. He is Fellow of IEEE and IET and is also on the editorial board of several international journals. He is the Editor-in-Chief for IEEE Wireless Communications Letters since 2020.
\end{IEEEbiography}

\begin{IEEEbiography}[{\includegraphics[width=1in,height=1.25in,clip,keepaspectratio]{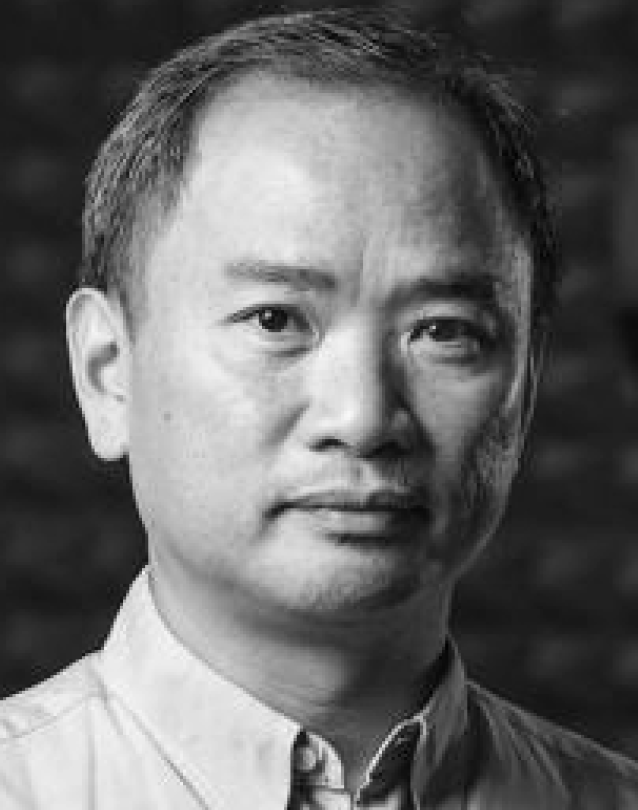}}]{Kin-Fai Tong} received the B.Eng. and Ph.D. degrees in electronic engineering from the City University of Hong Kong, Hong Kong, in 1993 and 1997, respectively. He was a Research Fellow with the Department of Electronic Engineering, City University of Hong Kong. He was the Post Expert Researcher with the Photonic Information Technology Group and Millimetre-Wave Devices Group, National Institute of Information and Communications Technology (NiCT), Tokyo, Japan, where his main research focused on photonic-millimeter-wave planar antennas for high-speed wireless communications systems. In 2005, he joined the Department of Electronic and Electrical Engineering, University College London (UCL), London, U.K., as a Lecturer, and where he is currently a Reader of antennas, microwave, and millimeter-wave engineering. His current research interests include millimeter-wave antennas, fluid antennas, 3-D printed antennas, and sub-gigahertz long-range Internet-of-Things (IoT) networks. Dr. Tong served as the General Co-Chair of the 2017 International Workshop on Electromagnetics (iWEM).
\end{IEEEbiography}

\begin{IEEEbiography}[{\includegraphics[width=1in,height=1.25in,clip,keepaspectratio]{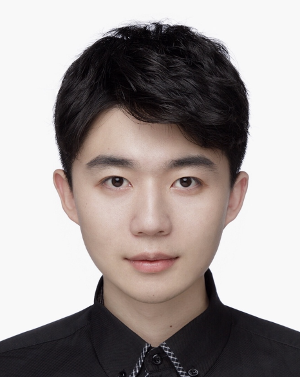}}]{Zhiyuan Chu} is a PhD student at the Department of Electronic and Electrical Engineering, University College London, UK, researching on the design and analysis of surface wave communications, and their application in 6G and computer processing systems. He received the BEng degree in Telecommunications Engineering with Management from Beijing University of Posts and Telecommunications, China, in 2017 and the MSc degree with Distinction in Electronic and Electrical Engineering from University of Sheffield, UK, in 2019.
\end{IEEEbiography}

\begin{IEEEbiography}[{\includegraphics[width=1in,height=1.25in,clip,keepaspectratio]{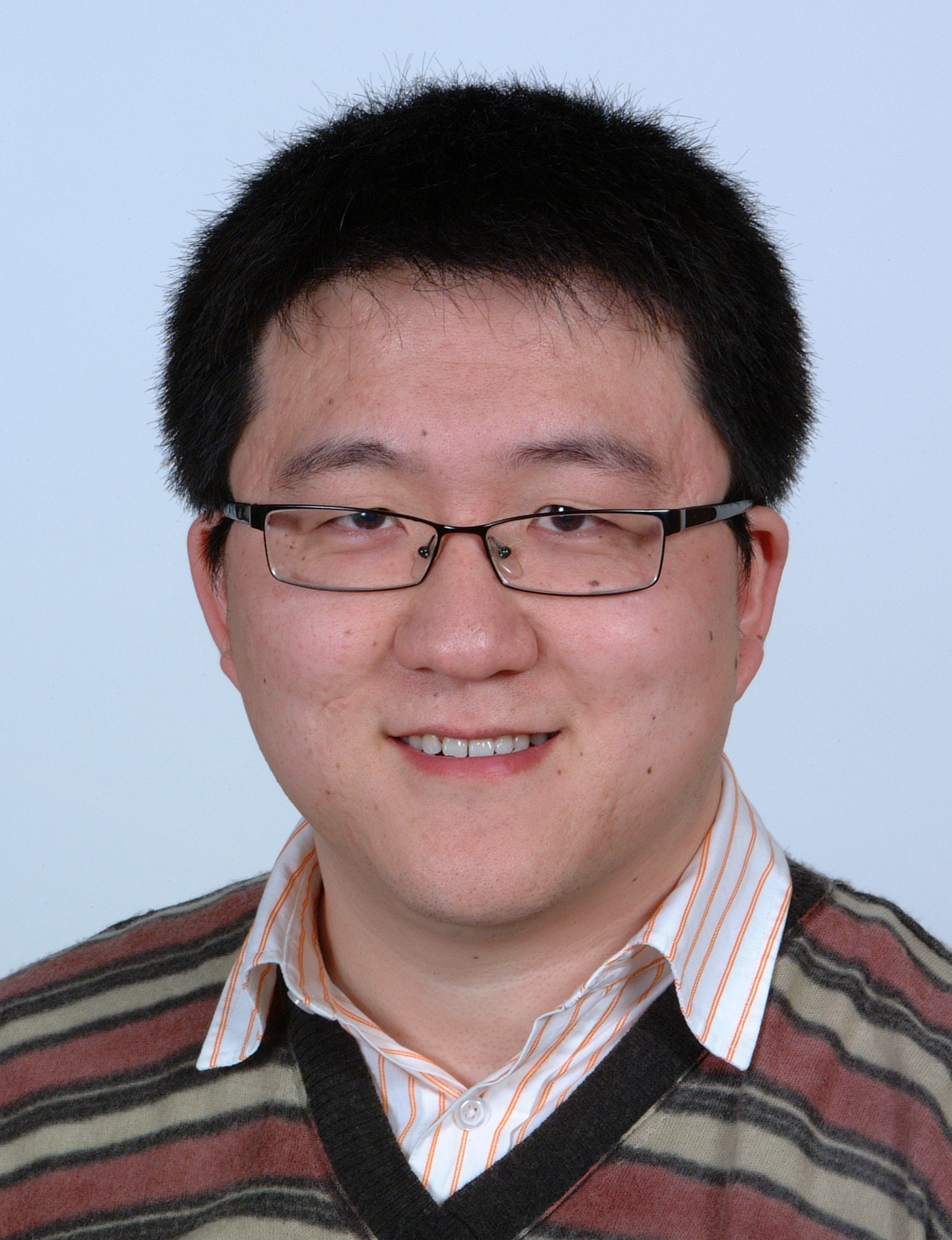}}]{Yangyang Zhang} received the B.S. and M.S. degrees in Electronics and Information Engineering from Northeastern University, Shenyang, China, in 2002 and 2004, respectively, and the Ph.D. degree in Electrical Engineering from the University of Oxford, Oxford, UK, in 2008. He is Executive Director at Kuang-Chi Institute of Advanced Technology, China. His research interests include multiple-input multiple-output wireless communications and stochastic optimization algorithms. Dr. Zhang has been awarded more than 20 honours. Besides, he also authored and co-authored more than 30 refereed papers. 
\end{IEEEbiography}




\end{document}